\documentclass[%
 reprint,
 amsmath,amssymb,
 aps,
 prb,
]{revtex4-2}

\usepackage{physics}
\usepackage{graphicx}
\usepackage{dcolumn}
\usepackage{bm}
\usepackage{hyperref}

\usepackage[normalem]{ulem} 

\hypersetup{citecolor=black,colorlinks=false,urlcolor=black}

\usepackage[utf8]{inputenc}
\usepackage[T1]{fontenc}
\usepackage{hyperref}
\usepackage{graphicx}
\usepackage{amsfonts, amsmath, amsthm, amssymb} 
\usepackage{braket}
\usepackage{float}
\usepackage{amsthm}
\usepackage{color}
\usepackage[all]{nowidow}

\hypersetup{citecolor=black,colorlinks=false,urlcolor=black} 
\usepackage[english]{babel}

\usepackage{csquotes}

\usepackage{xcolor}

\begin{document}


\title{Dynamical Crossover of the Quantum Fisher Information in the Spin-Boson Model}

\author{D. Parlato$^{1,\dag}$}\author{G. Di Bello$^{1}$}\author{F. Pavan$^{1}$}\author{G. De Filippis$^{2,3}$}\author{C. A. Perroni$^{2,3}$} 
\affiliation{$^{1}$Dip. di Fisica E. Pancini - Università di Napoli Federico II - I-80126 Napoli, Italy}
\affiliation{$^{2}$SPIN-CNR and Dip. di Fisica E. Pancini - Università di Napoli Federico II - I-80126 Napoli, Italy}
\affiliation{$^{3}$INFN, Sezione di Napoli - Complesso Universitario di Monte S. Angelo - I-80126 Napoli, Italy}
\affiliation{$^{\dag}$Corresponding author}

\begin{abstract}
We investigate the dynamical quantum Fisher information of a two-level system coupled to a bosonic environment, focusing on the estimation of the qubit gap. We combine analytical calculations with numerically controlled matrix-product-state simulations. In the exactly solvable pure-dephasing Ohmic regime at zero temperature, the long-time quantum Fisher information displays a coupling-dependent algebraic behavior, leading to a dynamical crossover: it grows without bound at weak coupling, approaches a finite asymptotic value at the crossover coupling, and vanishes at strong coupling. At finite temperature, thermal fluctuations suppress the long-time growth and generate a finite-time maximum, {whose dependence on the dephasing coupling retains a clear signature of the zero-temperature crossover.} We show that this crossover is absent at zero temperature in non-Ohmic baths: the long-time quantum Fisher information vanishes in the sub-Ohmic and diverges in the super-Ohmic regimes. {At non-zero temperature instead, the crossover appears for a super-Ohmic quadratic bath and the crossover coupling becomes temperature-dependent}. Moreover, in the 
zero-temperature Ohmic regime, we introduce an additional amplitude-damping system-bath coupling that induces energy relaxation. This 
relaxation channel replaces the unbounded long-time growth with a finite asymptotic quantum Fisher information associated with the reduced interacting ground state, {while a signature of the pure-dephasing crossover persists in the early-time dynamics.}
These results establish a direct connection between the low-frequency {structure of the bath} and the asymptotic metrological behavior of dynamical gap sensing, {and show how thermal fluctuations and energy relaxation regularize the ideal pure-dephasing dynamical crossover of the quantum Fisher information.}
\end{abstract}

\maketitle

\section{Introduction}\label{sec:introduction}
Quantum sensing has emerged as a promising field for the development of quantum technologies, with numerous present and future applications such as distributed sensing \cite{Zhang_2021,4vdx-7224,Kim2024,xia2026scalingenhancementdistributedquantum,rosati2026certificationnetworkquantumsensing,prabhu2026quantumcomputationaldisplacementsensing,Ho_2026}, critical sensing \cite{QFI_rabi,QFI_enh_sensing_2,Daniele,PRXQuantum.3.010354}, quantum computation \cite{Nielsen_Chuang_2010}, quantum thermometry \cite{winczewski2026nonequilibriumquantumthermometrybosonic}, among others. Although the main elements of the theory are well established, a gap still remains between theoretical models and their applicability, mainly due to noise and information loss \cite{Saleem,5gh9-nmv8}.
In this context, quantum sensing has developed along a research direction focusing on open systems as probes, since a quantum sensor inevitably interacts with its surrounding environment, leading to non-Markovian effects and, in some cases, to criticality-enhanced sensing \cite{adani2024critical,QFI_rabi,L__2022,Zhu_2023}. 

In this paper we study how the coupling to an environment affects the optimal sensitivity of a qubit, quantified by the quantum Fisher information (QFI) \cite{attainability,QFIM}. The QFI measures the sensitivity of a quantum state to variations of a parameter thanks to its link to the geometric structure of the parametric Hilbert space \cite{Sidhu_2020,referee3}.
Unlike the classical Fisher information $I(\rho;\Pi)$ which depends on the {probabilities of the outcomes of} a measurement $\Pi$, the QFI {$F(\rho)$} is measurement-independent and depends only on the probe state. This measurement-independence makes the QFI a natural tool for detecting criticality and optimizing probe states prior to the estimation process. A state with maximal QFI can, in principle, yield a more accurate estimate, although the precision ultimately achieved still depends on the measurement scheme employed to extract the parameter value. The relation between the theoretical and practically attainable precision is expressed by the quantum Cramér-Rao bound \cite{Nielsen2013,Alipour_2014}: 
\begin{equation}
\text{Cov}(\vec{{x}}) \ge \frac{1}{n} I({\rho(\vec{x}}),\Pi)^{-1} \ge \frac{1}{n} F(\rho(\vec{x}))^{-1}\;,\label{Eq_CRAMER_RAO}
\end{equation}
where
$\vec{x}=(x_1,\ldots,x_\mu,\ldots,x_\nu,\ldots)$ is a parameter vector and $\rho(\vec{x})$ is a quantum state depending on it. $\Pi$ is a positive operator-valued measure and $n$ is the number of repetitions of the measurement. In the single-parameter case it is straightforward to see that the {inverse} QFI sets a lower bound on the variance of an estimate of $x_\mu$.

Generally speaking, a sensing analysis from the QFI perspective aims to identify the probe state that maximizes the QFI, in the expectation that suitable experimental implementation can approach the corresponding bound \cite{Di-candia,ghosh2026journeyquantummetrologysensing}. The study of the ground-state QFI is particularly relevant for static sensors, for instance those exploiting enhanced sensitivity near quantum phase transitions. 
In the dynamical setting the QFI captures signatures of non-Markovianity: an increase in QFI (a positive flow) indicates information gain by the system, which is associated with information backflow \cite{QFI_non_markov,Breuer2,Lu_2010}. The time dependence of the QFI is therefore essential for understanding the effect of the environment on an open system. In particular, one typically finds that interactions modify the time dependence and reduce the Fisher information.\\ 

In this work, we investigate the dynamical behavior of the QFI in the spin-boson model (SBM), a paradigmatic open quantum system describing a qubit coupled to a set of $N$ harmonic oscillators representing the environment \cite{breuer2002theory, Hao_2013, Sun2025}.
We consider dephasing strengths ranging from weak to strong coupling and subsequently introduce an additional weak energy-relaxation channel. In a recent work \cite{Daniele}, we studied the QFI with respect to the system-bath coupling and its divergence near the quantum phase transition occurring in the strong-damping regime at zero temperature. The present work addresses a complementary physical setting: we focus on predominantly dephasing dynamics and estimate the qubit energy gap rather than the system-bath coupling. In particular, we study the QFI with respect to the qubit gap, investigating how dephasing, finite temperature, and energy relaxation affect the ultimate precision achievable in its estimation. Indeed, we show that the low-frequency structure of the environment determines a universal asymptotic behavior of the corresponding dynamical QFI.
To this aim, we combine analytical calculations with numerically controlled matrix-product-state (MPS) simulations.

Starting from the pure-dephasing case, we analytically show that, while an isolated system exhibits the quadratic time scaling 
\cite{Di-candia}, the coupling to the bath substantially modifies this behavior. In the zero-temperature Ohmic regime, we identify a dynamical crossover between weak and strong dephasing. At weak coupling, the QFI grows without bound in time, although with a reduced algebraic scaling compared with the uncoupled case, whereas at strong coupling it vanishes at long times. At the crossover coupling, it approaches a finite asymptotic value. This behavior provides a clear signature in the QFI, while conventional spin observables and the von Neumann entropy do not exhibit an equally pronounced distinction between the two regimes. We then extend the analytical study to finite temperature and show that thermal fluctuations suppress the long-time divergence. The QFI develops a finite-time maximum, and a signature of the zero-temperature crossover persists as a marked change in the curvature of this maximum as a function of the dephasing coupling. We expand this analysis to non-Ohmic environments. At zero temperature, we demonstrate that the QFI universally vanishes for sub-Ohmic baths, whereas super-Ohmic spectra restore the quadratic scaling characteristic of the uncoupled system. {The introduction of a non-zero temperature shifts the whole picture: the crossover appears for a super-Ohmic quadratic bath and the crossover coupling becomes temperature-dependent. }

Finally, we introduce an additional weak amplitude-damping coupling, which breaks the conservation of the qubit energy and induces relaxation. In the zero-temperature, weak-dephasing Ohmic regime, the QFI approaches at long times a finite value associated with the reduced interacting ground state, whereas a finite-time maximum and a signature of the crossover remain visible during the transient dynamics. 
Therefore, in the Ohmic case, finite temperature and amplitude damping provide two physically different mechanisms that regularize the asymptotic behavior. This establishes a unified framework for understanding dynamical gap sensing in structured bosonic environments.

The paper is organized as follows. {Section~\ref{sec:model_methods}} introduces the spin-boson model and the numerical methods. In the main text, we focus on the Ohmic spin-boson model. {Section~\ref{sec:puredeph}} focuses on the exactly solvable pure-dephasing Ohmic regime at zero temperature {(Subsec.~\ref{subsec:zeroT}), at finite temperature (Subsec.~\ref{subsec:finiteT}), {and an overview on the behavior for non-Ohmic environments (Subsec.~\ref{subsec:non-ohmic})}.
Section~\ref{sec:puredeph_weakdamp}} investigates the effect of an additional amplitude-damping coupling, {by analyzing both the statics (Subsec.~\ref{subsec:statics}) and the dynamics (Subsec.~\ref{subsec:dynamics})}. Appendix {\ref{app:algorithm}} details the numerical methods used in this work.

\section{Model and Numerical Methods}\label{sec:model_methods}
We adopt units where the Planck constant $\hbar=1$ and the Boltzmann constant $k_B=1$.
The Hamiltonian of the generalized SBM is written as \cite{Quantum_diss_syst}:
\begin{equation}H_{SB}=H_{S}+H_B+H_I\;, \label{Eq4}\end{equation}
where: $1)$ $H_S=-\frac{\Delta}{2}\sigma_x$ is the qubit free Hamiltonian, with $\Delta$ representing the qubit gap; $2)$ $H_B=\sum_{i=1}^N\omega_i\bigg(b_i^\dagger b_i+\frac{1}{2}\bigg)$ is the bath Hamiltonian, describing a set of $N$ harmonic oscillators with frequencies $\omega_i$; $3)$ $H_I=\sum_{i=1}^N(\sigma_x\lambda_i^x+ \sigma_z\lambda_i^z)(b_i+b_i^\dagger)$ is the qubit-bath interaction Hamiltonian, with $\lambda_i^x$ and $\lambda_i^z$ representing the longitudinal and transverse couplings, respectively. Here, $\sigma_z$ and $\sigma_x$ are the Pauli matrices representing the two-level system, and $b_i$ and $b_i^\dagger$ are the annihilation and creation operators for bosonic modes of frequency $\omega_i$. The couplings $\lambda_i^x$ are obtained by discretizing a continuous spectral density function $J_x(\omega)=\sum_i^N|\lambda_i^x|^2\delta(\omega-\omega_i)=\kappa f_s(\omega)$, with $f_s(\omega)= \frac{1}{2}\omega^s\omega_c^{1-s}e^{-\frac{\omega}{\omega_c}}$. Here $\omega_c$ is the cutoff frequency that we set to be our energy unit, $\kappa$ is the dimensionless spin-bath coupling associated with pure-dephasing processes; $s$ is the ohmicity parameter ($s=1$ for Ohmic environments). The couplings $\lambda_i^z$ are sampled analogously and we denote the dimensionless strength of amplitude-damping processes by $\alpha$; therefore, $J_z(\omega)=\alpha f_s(\omega)$.
\\

The quantum Fisher information matrix (QFIM) can be calculated using the spectral decomposition of the density operator $\rho$: 
\begin{equation}F_{\mu\nu}=\sum_{i,j}\frac{2Re(\bra{\lambda_i}\partial_\mu\rho\ket{\lambda_j}\bra{\lambda_j}\partial_\nu\rho\ket{\lambda_i})}{\lambda_i+\lambda_j},\label{Eq2}\end{equation}
where $\lambda_i$ and $\ket{\lambda_i}$ are, respectively, eigenvalues and eigenvectors of $\rho$, and $\partial_\mu$ is the derivative of the state with respect to the parameter $\mu$.
A useful expression is available for the particular case of one qubit \cite{QFIM}:
\begin{equation}F_{\mu\nu}=Tr[(\partial_\mu\rho)(\partial_\nu\rho)]+\frac{Tr[\rho(\partial_\mu\rho)\rho(\partial_\nu\rho)]}{\det \rho},\label{Eq3}\end{equation}
where $\det(\rho)$ is the determinant of $\rho$.

We compute the Hamiltonian ground state (GS) and the dynamics of the QFIM elements using analytical approximations as well as numerical techniques. Analytical treatments of the dynamics in the case of pure dephasing rely 
on the exact solution available for this model \cite{breuer2002theory,Shaller}. 
However, the addition of amplitude damping requires MPS-based numerical methods \cite{grazia1,grazia2,grazia3}. Indeed,
tensor-network techniques based on the MPS representation yield numerically controlled results for models mapped onto one-dimensional chains, using the ITensor library \cite{fishman2022itensor}. The GS has been calculated using the density-matrix renormalization group (DMRG) with a cutoff of $10^{-15}$ \cite{schollwock2011density}. The dynamics have been simulated 
using the TEMPO algorithm \cite{Strathearn2018}, as implemented in the OQuPy library \cite{8fbc5c70103941f79cd049122c3ff946}. {To validate the long-time dynamics, we also checked the results using the TDVP algorithm} \cite{paeckel2019time}, which we do not show here. We refer {the reader} to the Appendix {\ref{app:algorithm}} for an extended explanation {of our choice of algorithm for the dynamics.} \\

\section{Analytical QFI of pure dephasing}\label{sec:puredeph}
We first set $\alpha=0$, so that the Hamiltonian reduces to its well-known pure-dephasing form. {Since $[H,H_S]=0$, the qubit energy operator, proportional to $\sigma_x$, is a constant of motion.} We assume {an initial} factorized state $\rho_{tot}({0})=\rho({0})\otimes\rho_B$, where $\rho_B$ is the thermal state of the bath. The dynamics can be obtained analytically and in the $\sigma_x$ eigenbasis are given by \cite{breuer2002theory}:
\begin{equation}
\rho(t)=\bigg(\begin{array}{cc}
\rho_{00}(0) & \rho_{01}(0)e^{-\kappa\Gamma(t)+i\Delta t}\\
\rho_{10}(0)e^{-\kappa\Gamma(t)-i\Delta t} & \rho_{11}(0)
\end{array}\bigg),\label{Eq5}
\end{equation}
{The initial state is chosen as} 
\begin{equation}
\rho(0)=\left(\begin{array}{cc}
\frac{1}{2} & \frac{1}{2}\\
\frac{1}{2} & \frac{1}{2}
\end{array}\right)\;,\label{Eq5b}
\end{equation}
{and will be such throughout the whole paper.}
$\Gamma(t)$ is the decoherence function, that is calculated as
\begin{equation}\Gamma(t)=\int_0^{+\infty}d\omega\; \frac{J(\omega)}{\kappa}\coth\bigg(\frac{\beta\omega}{2}\bigg)\frac{1-\cos\omega t}{\omega^2}\;.\end{equation}
We note that the division by $\kappa$ is only done to make explicit the dependence of the state on the coupling, which will be useful for the calculations. Assuming $s=1$, the decoherence function is given by:
\begin{equation}\Gamma(t)
 =\ln(1+\omega_c^2 t^2)+\ln\bigg(\frac{\sinh(\pi t/\beta)}{\pi t/\beta }\bigg)\;.\end{equation} 
 In Subsec.~(\ref{subsec:non-ohmic}) we also consider non-Ohmic spectral densities. The conservation of the qubit energy is reflected in the fact that the populations $\rho_{ii}$ do not evolve in time; as a consequence, there is no relaxation, but only dephasing. 

\subsection{Zero temperature}\label{subsec:zeroT}
\begin{figure}[H]
    \centering
    \includegraphics[width=0.8\linewidth]{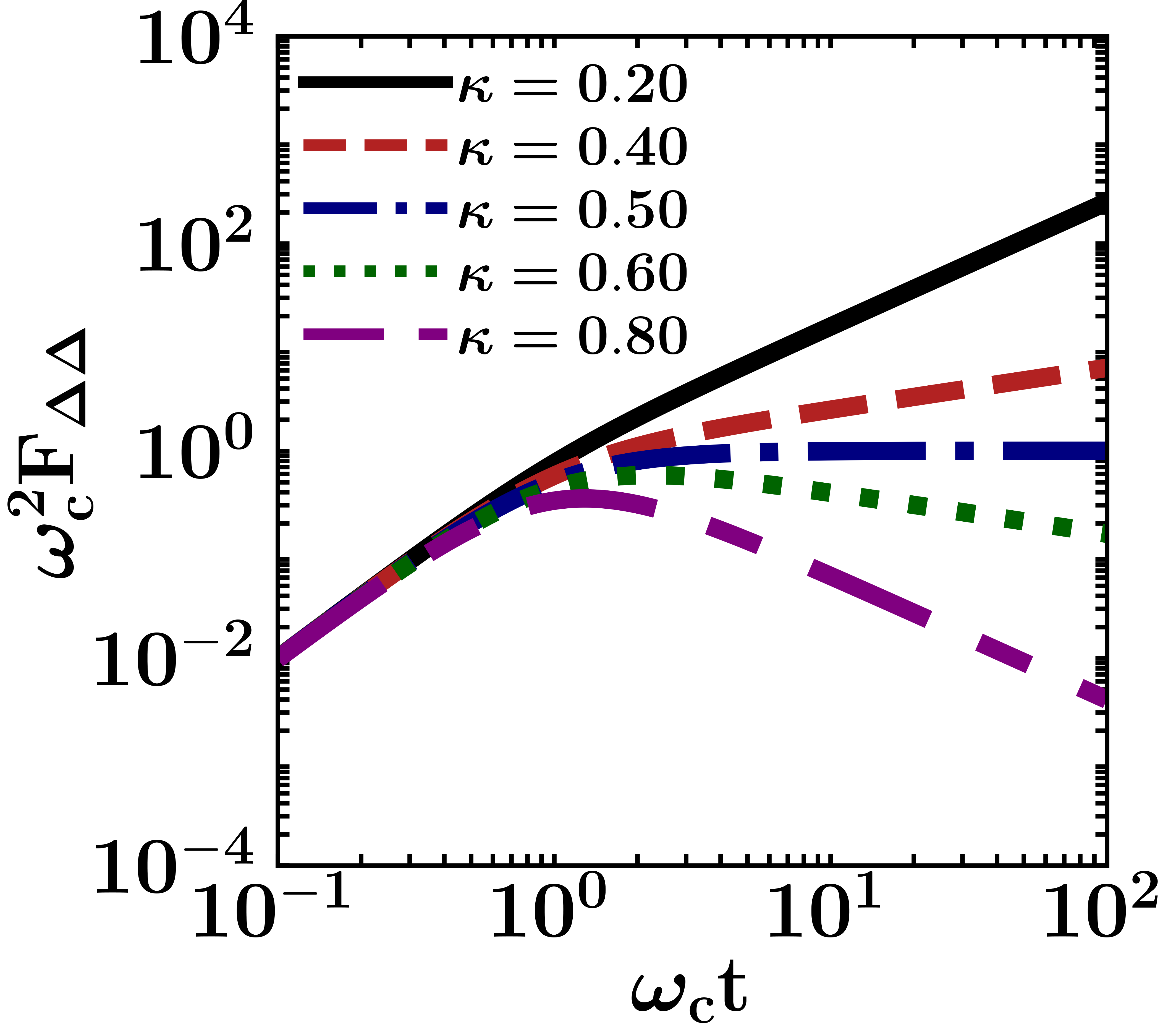}
    \caption{QFI for the qubit gap $F_{\Delta\Delta}$ as a function of time for $T=0$, $\alpha=0$, $s=1$, and various dephasing couplings $\kappa$ (left), and $F_{\Delta\Delta}(t)$ for a fixed coupling $\kappa=0.5$ and different ohmicity values $s$ (right)}.
    \label{Fig1}
\end{figure}

We now investigate how the sensitivity of the state to $\Delta$ evolves at $T=0$. Given the analytical form of Eq.~\eqref{Eq5}, the QFI $F_{\Delta\Delta}$ can be directly computed using Eq.~\eqref{Eq3}:
\begin{equation}
F_{\Delta\Delta}=4|\rho_{01}(0)|^2 t^2 e^{-2\kappa\Gamma(t)}
=\frac{t^2}{(1+\omega_c^2 t^2)^{2\kappa}}.
\label{Eq6}\end{equation}
{The resulting QFI is shown} in the left panel of Fig.~(\ref{Fig1}) for various $\kappa$ {and} $s=1$.
Since at zero temperature the decoherence function scales only logarithmically, the QFI follows a power law with a $\kappa$-dependent exponent $F_{\Delta\Delta}\sim t^{2(1-2\kappa)}$. This leads to a crossover at $\kappa=1/2$ between two distinct regimes: for $\kappa<1/2$ (weak-dephasing) the QFI diverges in time; for $\kappa>1/2$ (strong-dephasing), it vanishes asymptotically; at $\kappa=1/2$ the QFI goes to the finite value $F_{\Delta\Delta}\rightarrow1/\omega_c^2$. 
The physical mechanism behind this behavior is the competition between coherence loss induced by the environment, captured by the exponential decay, and the phase accumulation associated with the qubit free precession, represented by the $t^2$ term. If the qubit is prepared in a diagonal state $\rho_{01}(0)=\rho_{10}(0)=0$, the QFI is always null, since the gap dependence is only in the coherence. In general, the decoherence term follows a pure exponential decay, which would always dominate at long times; however, at zero temperature its power-law dependence on time allows it to compete with the phase accumulation term, rather than suppress it outright. We note that this crossover cannot be found in any other {analyzed} quantity related to the qubit dynamics, in particular we can see that {the density matrix depends smoothly on} $\kappa$. \\
An important aspect is the attainability of the bound set by the Fisher information, which depends on the choice of measurement. Estimating the gap $\Delta$ requires measuring some observable sensitive to it, raising the question of which observable enables optimal estimation. This is addressed by studying the signal{-to-}noise ratio (SNR) of a given observable $\hat{O}$, defined as \cite{Di-candia}:
\begin{equation} SNR_\Delta(O)=\frac{|\partial_\Delta\langle\hat{O}\rangle|^2}{\langle\hat{O}^2\rangle-\langle\hat{O}\rangle^2}\;.\label{Eq10}\end{equation}
Since the Bloch-vector dynamics are confined to the $yz$ plane, {the observable that best tracks changes in $\Delta$ is the one aligned with the direction along which the Bloch vector moves as $\Delta$ changes, so tangent to its trajectory and thus orthogonal to the Bloch vector itself within that plane.} Indeed, the operator:
\begin{equation}\hat{O}_{opt}=\sin(\Delta t)\sigma_z-\cos(\Delta t)\sigma_y\;,\label{Eq11}\end{equation}
has a SNR equal to the QFI, so that measuring $\hat{O}_{opt}$ saturates the Cramér-Rao bound and provides the best possible sensitivity to $\Delta$. Notably, at the specific times $\Delta t^*=\pi/2+2\pi m$, $\hat{O}_{opt}$ reduces to $\sigma_z$ alone, so that optimal estimation can be achieved with the standard Ramsey protocol rather than the more demanding joint measurement of both observables. Thus, in the weak-dephasing regime at $T=0$, when the interrogation time itself is not treated as a constrained resource, the single-shot QFI can be increased by extending the interrogation time, provided that the measurement is performed at the optimal times $\Delta t^*$ defined above. In realistic scenarios with $T\neq0$, a trade-off becomes necessary due to the unavoidable decay of sensitivity at long times, as will be discussed in Subsec.~(\ref{subsec:finiteT}). {We note, however, that this process presupposes prior knowledge of $\Delta$ itself, which the optimal measurement then helps to refine.}

{
To illustrate concretely how this prior knowledge enters the estimation, consider estimating $\Delta$ using a Ramsey protocol. The QFI quantifies local sensitivity and therefore presupposes that an approximate value of the parameter, $\Tilde{\Delta}_0\pm\delta_0$, is already known. In the Ramsey scheme, one can measure $\langle\sigma_z\rangle=e^{-\kappa\Gamma(t)}\cos(\Delta t)$ at a fixed interrogation time $t_0$. Since the cosine is periodic in $\Delta$ with period $2\pi/t_0$, the signal shows the well-known Ramsey fringes. From a metrological perspective, this means that if $t_0>2\pi/\delta_0$, more than one fringe falls within the initial error interval, making it impossible to distinguish $\Delta$ from $\Delta+2\pi m/t_0$ ($m\in \mathbb{N}$). This ambiguity is exacerbated as $t_0\rightarrow +\infty$, which is exactly the limit one would like to approach to optimize sensitivity in the weak-coupling regime. The way out is to adaptively optimize the interrogation time, keeping it short enough that only a single fringe lies within the current uncertainty interval. A first one-shot measurement at time $t_0$, when the Cramér-Rao bound is saturated, yields an estimate with an uncertainty $\delta_1=1/\sqrt{F(t_0)}$ ($n=1$ in Eq.~\ref{Eq_CRAMER_RAO}). The interrogation time can then be increased up to a value $t_1<2\pi/\delta_1$, and the measurement repeated. Iterating this procedure progressively reduces the estimation error, thanks to the unbounded QFI growth. This strategy, however, cannot be continued in the strong-coupling regime, where a well-defined optimal interrogation time exists, beyond which the QFI decreases and the uncertainty increases again.}

\subsection{Finite temperature}\label{subsec:finiteT}
\begin{figure}
    \centering
    \includegraphics[width=0.8\linewidth]{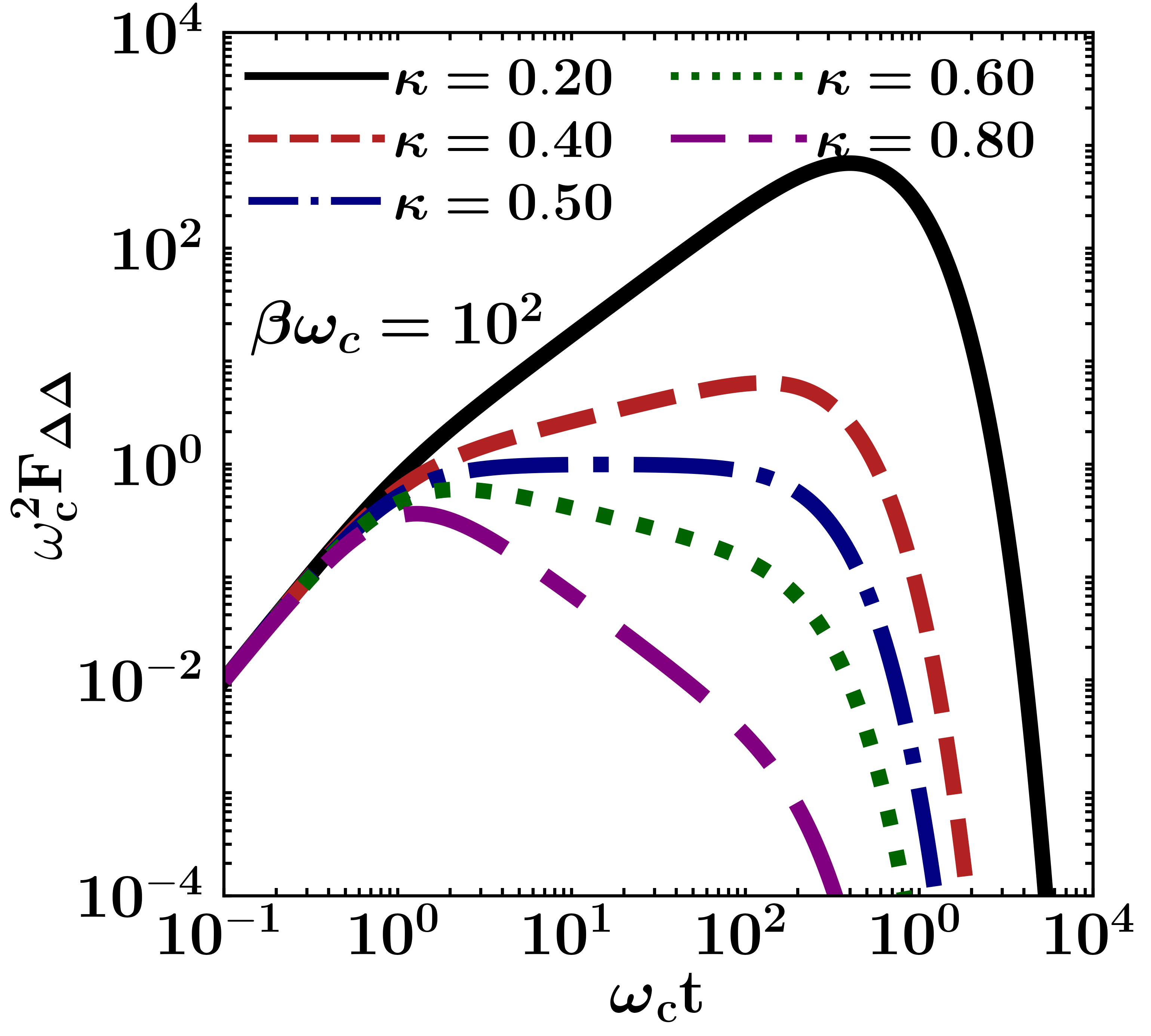}
    \caption{QFI for the qubit gap $F_{\Delta\Delta}$ as a function of time for $\beta=1/T=100$, $\alpha=0$, and various dephasing couplings $\kappa$.}
    \label{Fig2}
\end{figure}

At $T\neq 0$, the additional contribution to the decoherence function introduces a stronger suppression that cannot be compensated by phase accumulation, so that $F_{\Delta\Delta}\rightarrow 0$ for any $\kappa$:
\begin{equation}
F_{\Delta\Delta}=4|\rho_{01}(0)|^2\frac{t^{2}}{(1+\omega_c^2 t^2)^{2\kappa}}\bigg(\frac{\sinh(\pi t/\beta )}{\pi t/\beta }\bigg)^{-2\kappa},
\label{Eq7}\end{equation}
shown in Fig.~(\ref{Fig2}).

\begin{figure}
    \centering
    \includegraphics[width=0.8\linewidth]{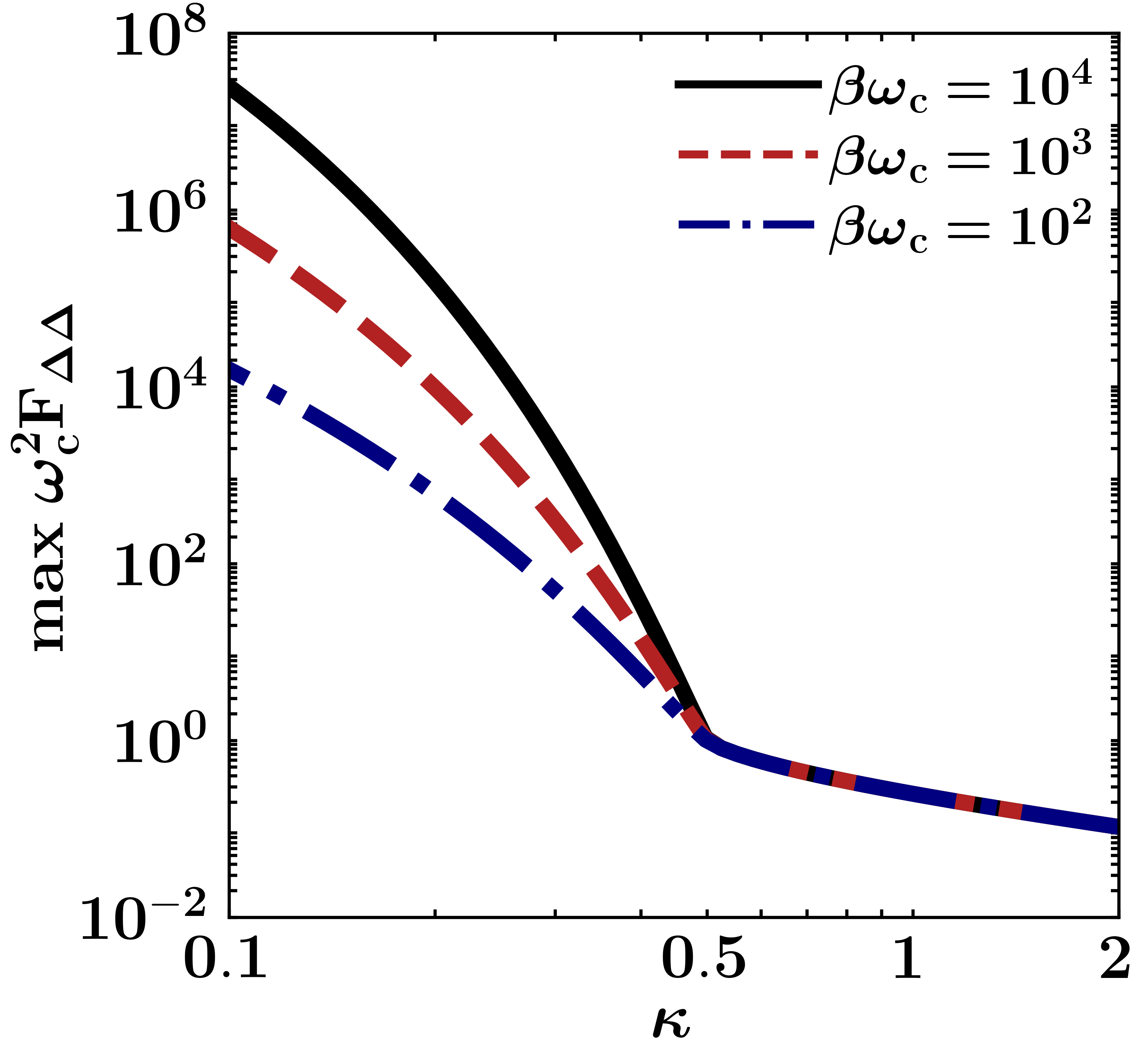}
    \caption{Maximum of QFI for the qubit gap $F_{\Delta\Delta}$ as a function of dephasing coupling $\kappa$ for $\alpha=0$ and several $T\neq0$.}
    \label{Fig3}
\end{figure}

Although the asymptotic divergence disappears, the crossover remains visible in the behavior of the maximum of $F_{\Delta\Delta}$ over time, whose curvature changes abruptly at $\kappa=1/2$ as is shown in Fig.~(\ref{Fig3}). In the limit $T\rightarrow 0$ ($\beta\rightarrow+\infty$), the left branch of the curve continues to grow without bound, whereas the right branch converges to the $T=0$ values. This confirms that the effect is driven by quantum fluctuations, which remain strong enough to persist even in the low-temperature regime.
To understand this behavior we look at the time dependence of Eq.~\eqref{Eq7} for times shorter than the thermal relaxation time $t/\beta\ll1$: 
\begin{equation*}
F_{\Delta\Delta}=4|\rho_{01}(0)|^2\frac{t^{2}}{(1+\omega_c^2 t^2)^{2\kappa}},
\end{equation*}
since $\sinh(x)/x\sim 1$ for small $x$. As expected, for short times the system sees the environment as a zero-temperature reservoir, so the crossover behavior reappears in the transient dynamics. The main difference here is that once the system starts to feel the effect of the bath ($t/\beta\approx1$) the QFI $F_{\Delta\Delta}$ exhibits an exponential decay and goes to zero. To summarize: for $\kappa>1/2$ the transient QFI reaches approximately the same maximum as in the $T=0$ case, and {then decays exponentially to zero.} For $\kappa<1/2$, the QFI initially grows and continues to do so until $t/\beta\approx1$, after which it starts decreasing. While it is true that it does not diverge, the nature of its maximum is very different from the one obtained for $\kappa>1/2$, and heavily depends on $\beta$: a larger $\beta$ requires a larger $t$ to produce the same effect, and thus the QFI {continues to grow over a longer time interval.}

The abrupt change appears also in the position of the maximum $\bar{t}$, which can indeed be seen to jump from long to short times as $\kappa$ {exceeds} $\kappa=0.5$. We can formally obtain $\bar{t}$ by solving:
\begin{equation}\frac{\bar{t}^2}{1+\bar{t}^2}+\frac{\pi \bar{t}}{\beta}\coth{\bigg(\frac{\pi\bar{t}}{\beta}\bigg)}-1=\frac{2}{\kappa}\;.\label{Eq8}\end{equation}
With suitable approximations, the solutions are found to be: \begin{equation}\bar{t}=\bigg\{\begin{array}{cc}\frac{\beta}{2\pi\kappa} & \kappa\ll1/2\\ \bigg(\frac{1}{6}(A(\kappa)+\sqrt{A(\kappa)^2+B(\kappa)})\bigg)^{-1/2} & \kappa\gtrsim1/2\end{array}\;\label{Eq9} \;,\end{equation}
where $A(\kappa)=\frac{2\pi^{2}\kappa}{\beta^{2}}+3\left(2\kappa-1\right)$ and $B(\kappa)=\frac{24\pi^{2}\kappa}{\beta^{2}}$. Within this approximation, the abrupt crossover disappears, since the solution is continuous across $\kappa \sim1/2$.
Taking the limit $\beta\rightarrow\infty$ shows that $\bar{t}\rightarrow\infty$ for $\kappa<1/2$: the maximum is pushed to infinite times, so that the divergence is recovered.
 \\ \\

\subsection{Non-Ohmic environments}\label{subsec:non-ohmic}
\begin{figure}
    \centering
    \includegraphics[width=1\linewidth]{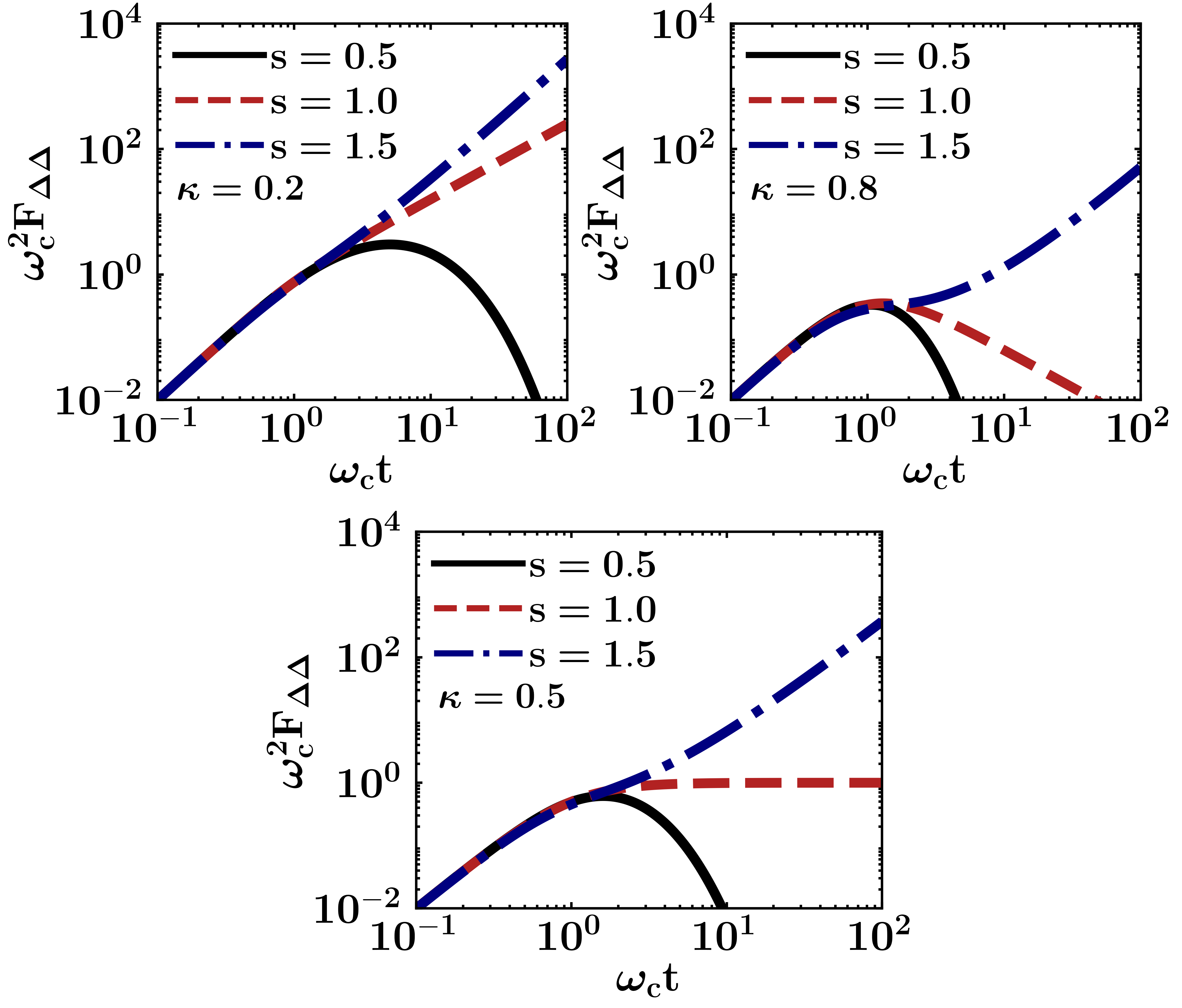}
    \caption{Dynamic quantum Fisher information $F_{\Delta\Delta}$ for three dephasing couplings $\kappa=0.2,0.5,0.8$ and three ohmicity parameters $s=0.5,1,1.5$. }
    \label{Fig6}
\end{figure}

We extend our study to general spectral densities to determine whether the pure-dephasing crossover in the QFI dynamical behavior is unique to the Ohmic regime. 
To this end, we recall the decoherence function from Eq.~\eqref{Eq5}. While the analytical structure of the reduced density matrix $\rho(t)$ stays the same, for an arbitrary ohmicity $s$, at zero temperature, the decoherence function takes the following form:
\begin{equation*} \Gamma(t)=\Gamma^{e}(s-1)\bigg[1-\frac{\cos((s-1)\arctan(\omega_ct))}{(1+\omega_c^2t^2)^{(s-1)/2}}\bigg]\;,
\end{equation*}
where $\Gamma^e$ is the Euler gamma function. Given that $\rho(t)$ retains the form in Eq.~\eqref{Eq5}, the QFI can again be analytically calculated as in Eq.~\eqref{Eq6}, with this generalized decoherence function. Fig.~(\ref{Fig6}) displays the temporal evolution of the QFI for several values of the coupling $\kappa$. 
Notably, the crossover behavior is absent in the non-Ohmic regimes. Specifically, a sub-Ohmic bath leads to a decaying QFI, whereas the super-Ohmic case results in a diverging QFI regardless of $\kappa$.

\begin{figure}
    \centering
    \includegraphics[width=1\linewidth]{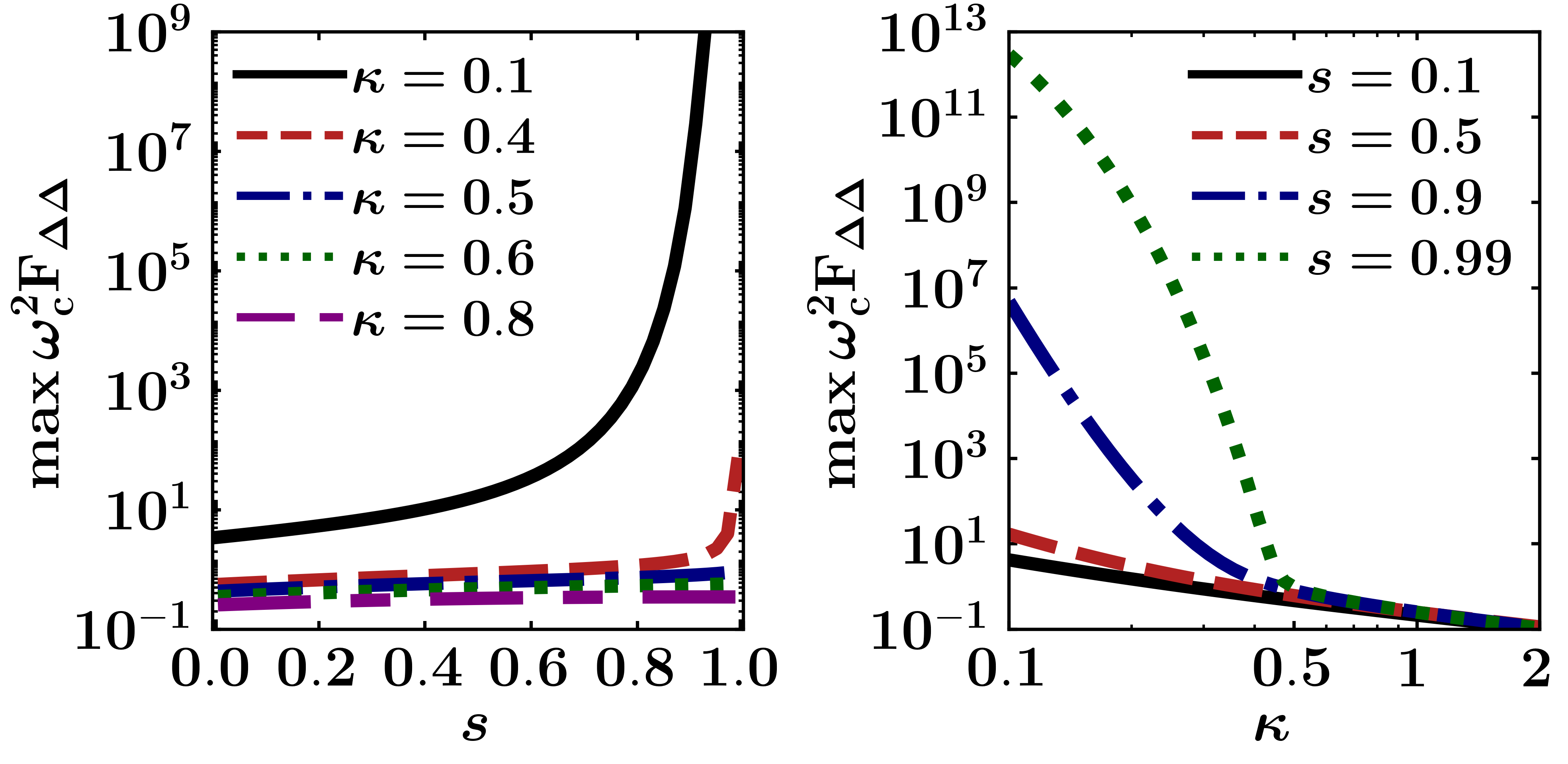}
    \caption{Maximum in time of the quantum Fisher information $\underset{t}{\max} F_{\Delta\Delta}(t)$ varying the ohmicity $s$ for fixed couplings $\kappa$ (left) and vice versa (right).}
    \label{Fig7}
\end{figure}

To determine whether the QFI crossover in $\kappa$ is preserved, we compute the long-time limits of $\Gamma(t)$ and $F_{\Delta\Delta}$ for a generic $s$.
The decoherence function yields:
\begin{equation}\Gamma(t)\approx\begin{cases}-\Gamma^e(s-1){\cos(\frac{\pi}{2}(s-1))}{(1+\omega_c^2t^2)^{(1-s)/2}} & s<1 \\ \ln(1+\omega_c^2t^2) & s=1 \\ \Gamma^e(s-1) & s>1\end{cases}\;.\end{equation}
Evidently, the Ohmic regime is peculiar, being the sole case where decoherence scales logarithmically with time.
In the sub-Ohmic regime, the power-law dependence of the decoherence function leads to an exponential decay of the Fisher information for any coupling $\kappa$. Conversely, the super-Ohmic regime yields only partial decoherence. After the initial bath interaction saturates, phase accumulation remains unsuppressed, restoring the quadratic QFI scaling typical of isolated dynamics:
\begin{equation}
F_{\Delta\Delta}(t)\overset{t\rightarrow+\infty}{\longrightarrow}
\begin{cases}
0 & s<1\\
0 \ (\kappa<\tfrac12),\;\; \frac{1}{\omega_c^2} \ (\kappa=\tfrac12),\;\; +\infty \ (\kappa>\tfrac12) & s=1\\
+\infty & s>1
\end{cases}
\end{equation}
Thus, the QFI crossover depends strictly on the linear behavior of the low-frequency spectral density. An enhancement of low-frequency modes accelerates information loss, whereas their suppression limits environmental noise, allowing phase accumulation to dominate. Focusing on the sub-Ohmic regime, which, {unlike the super-Ohmic regime,} exhibits a global maximum, Fig.~(\ref{Fig7}) shows the QFI maximum for varying $s$ at fixed $\kappa$ (left) and for varying $\kappa$ at fixed $s$ (right). As $s\rightarrow1^-$, weak-coupling $\kappa<1/2$ drives the maximum to divergence, while strong coupling yields a finite asymptote, consistently with the results of the Ohmic limit. The right panel of Fig.~(\ref{Fig7}) illustrates the same trend from a different {perspective}: the sub-Ohmic peak QFI grows within the weak-coupling regime as $s$ reaches $1$.

\begin{figure}
    \centering
    \includegraphics[width=1\linewidth]{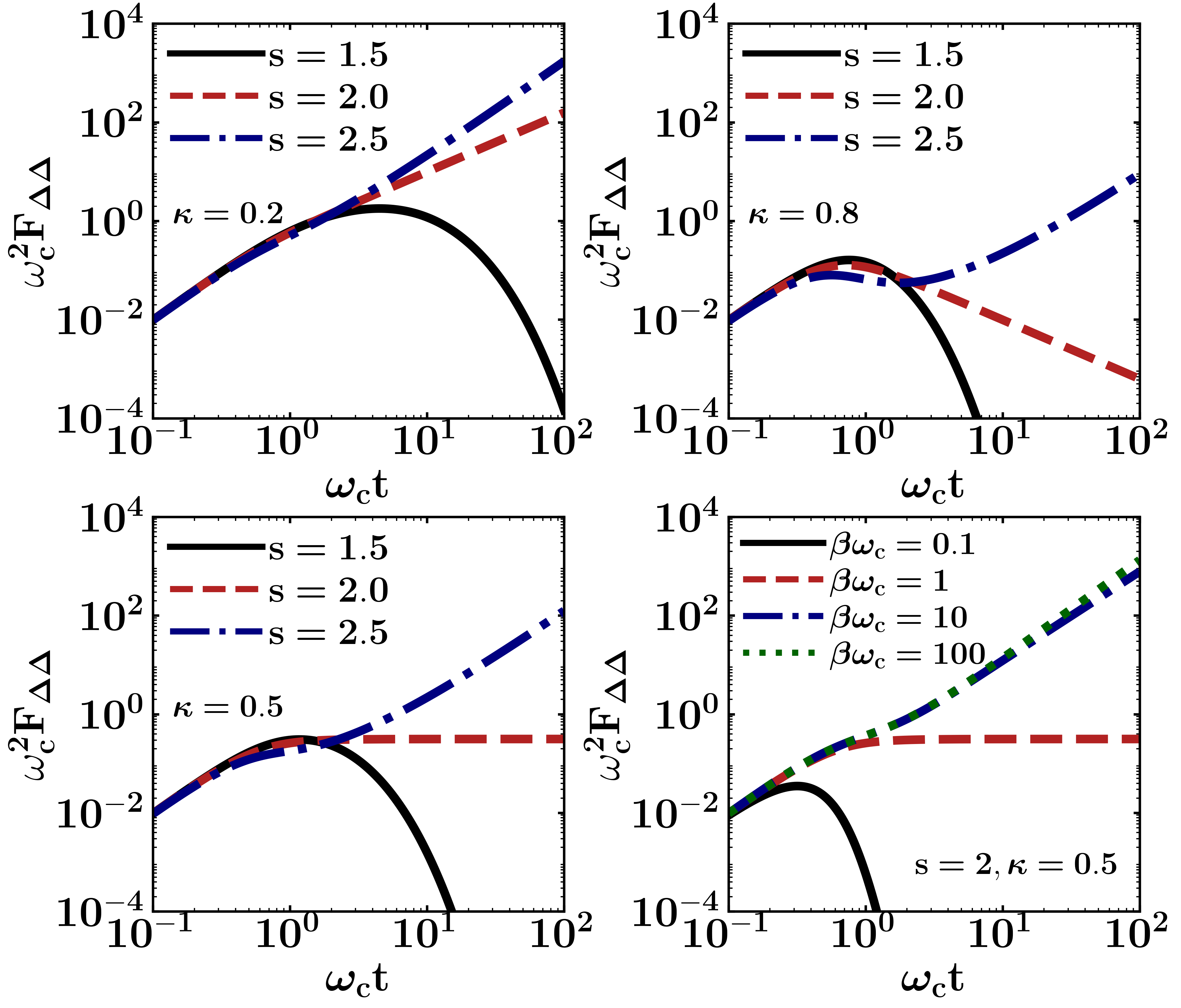}
    \caption{Dynamic quantum Fisher information $F_{\Delta\Delta}$ for three dephasing couplings $\kappa=0.2$ (top left) $\kappa=0.8$ (top right) and $\kappa=0.5$ (bottom left) and three ohmicity parameters $s=1.5,2,2.5$ for a fixed temperature $\beta\omega_c=100$. The bottom right panel shows $F_{\Delta\Delta}$ at $\kappa=0.5$ and $s=2$ for different $\beta\omega_c$.}
    \label{Fig6_2}
\end{figure}

We also extend our study to non-Ohmic environments at nonzero temperature. Accounting for a finite $\beta$ yields {the following decoherence function}:
\begin{gather}
\Gamma_T(t,s) = \Gamma_{T=0}(t,s) \notag\\
+ 2\,\Gamma^e(s-1)\sum_{n=1}^{\infty}\left\{a_n^{1-s} - b_n(t)^{1-s}\cos\!\left[(s-1)\,\theta_n(t)\right]\right\}
\end{gather}

where $a_n=\frac{1}{\omega_c}+\beta n$ encodes the temperature dependence, $b_n(t) \equiv \left[a_n^2+t^2\right]^{1/2}$ and $
\theta_n(t) \equiv \arctan(t/a_n)$. While this closed analytical form is convenient for numerical evaluations, extracting the asymptotic QFI behavior directly from it is challenging. To gain insight into this limit, we return to the general integral representation of the decoherence function for arbitrary $s$ and $\beta$:
\begin{equation}\Gamma(t)=\int_0^{+\infty}d\omega\; \frac{1}{2}\omega^s\omega_c^{1-s}e^{-\omega/\omega_c}\coth\bigg(\frac{\beta\omega}{2}\bigg)\frac{1-\cos\omega t}{\omega^2}\;,\end{equation}
In particular, assuming a fixed temperature $\beta$ and long evolution times $\omega t\gg1$ implies that the main contribution to the integral comes from the low frequencies $\beta\omega\ll1$. This finally allows us to approximate the hyperbolic cotangent term, which gives us:
\begin{equation}\Gamma(t)=\int_0^{+\infty}d\omega\; \omega^{s-1}\omega_c^{2-s}e^{-\omega/\omega_c}\bigg(\frac{1}{\beta\omega_c}\bigg)\frac{1-\cos\omega t}{\omega^2}\;.\end{equation}
This result indicates that the long-time behavior of $\Gamma(t)$ at finite temperature is equivalent to the zero-temperature asymptotic behavior of a bath with an effective ohmicity reduced by one, $s_{\mathrm{eff}}=s-1$. Consequently, for $s>2$ partial decoherence is restored (which at $T=0$ we had at $s>1$) and the QFI diverges; for $s<2$ the system undergoes complete decoherence and the QFI vanishes. A signature of the crossover is preserved at $s=2$ (the finite-temperature counterpart of $s=1$ at zero temperature), where $\Gamma(t)$ exhibits a logarithmic asymptotic growth, once again giving rise to the characteristic QFI crossover. {These different behaviors are depicted in Fig.~(\ref{Fig6_2}). The bottom right panel shows that the crossover coupling depends on the inverse temperature $\beta\omega_c$. In particular, since the long-time QFI scaling becomes $F_{\Delta\Delta}\sim t^{2-4\frac{\kappa}{\beta\omega_c}}$, the crossover coupling is $\overline{\kappa}=\beta\omega_c/2$. This implies that for a lower $\beta$ (higher temperature) the crossover appears for a lower coupling: a weaker dephasing is required for $F_{\Delta\Delta}$ to vanish due to the effect of temperature.}

\section{QFI with pure dephasing and weak damping}\label{sec:puredeph_weakdamp}
\subsection{Statics}\label{subsec:statics}
To better understand the origin of this crossover, we also analyze the QFI obtained from the Hamiltonian in the presence of a weak amplitude-damping term $\alpha$. {For both the static and dynamics we set $\Delta=\omega_c/10$}. 
We start by studying the QFI of the ground state, which in this case is nontrivial. Indeed, in the pure dephasing case $\alpha=0$, the GS of the qubit is $\ket{+}_z$ (because of the $-$ sign in $H_S$), independently from the $\Delta$ value, so that the static QFI is zero $F_{\Delta\Delta}=0$ for any value of the coupling $\kappa$. 
We note that this is not {inconsistent} with the existence of the crossover, since the pure dephasing model does not show relaxation, so that the stationary state (the one exhibiting the crossover) is not actually the ground state.
When turning on the damping interaction $\alpha\neq0$, the system relaxes to the ground state that becomes the stationary state of the qubit. Moreover, in this nontrivial case the derivative $\partial_\Delta\rho$ is nonzero and so is the QFI.
\begin{figure}
    \centering
    \includegraphics[width=1\linewidth]{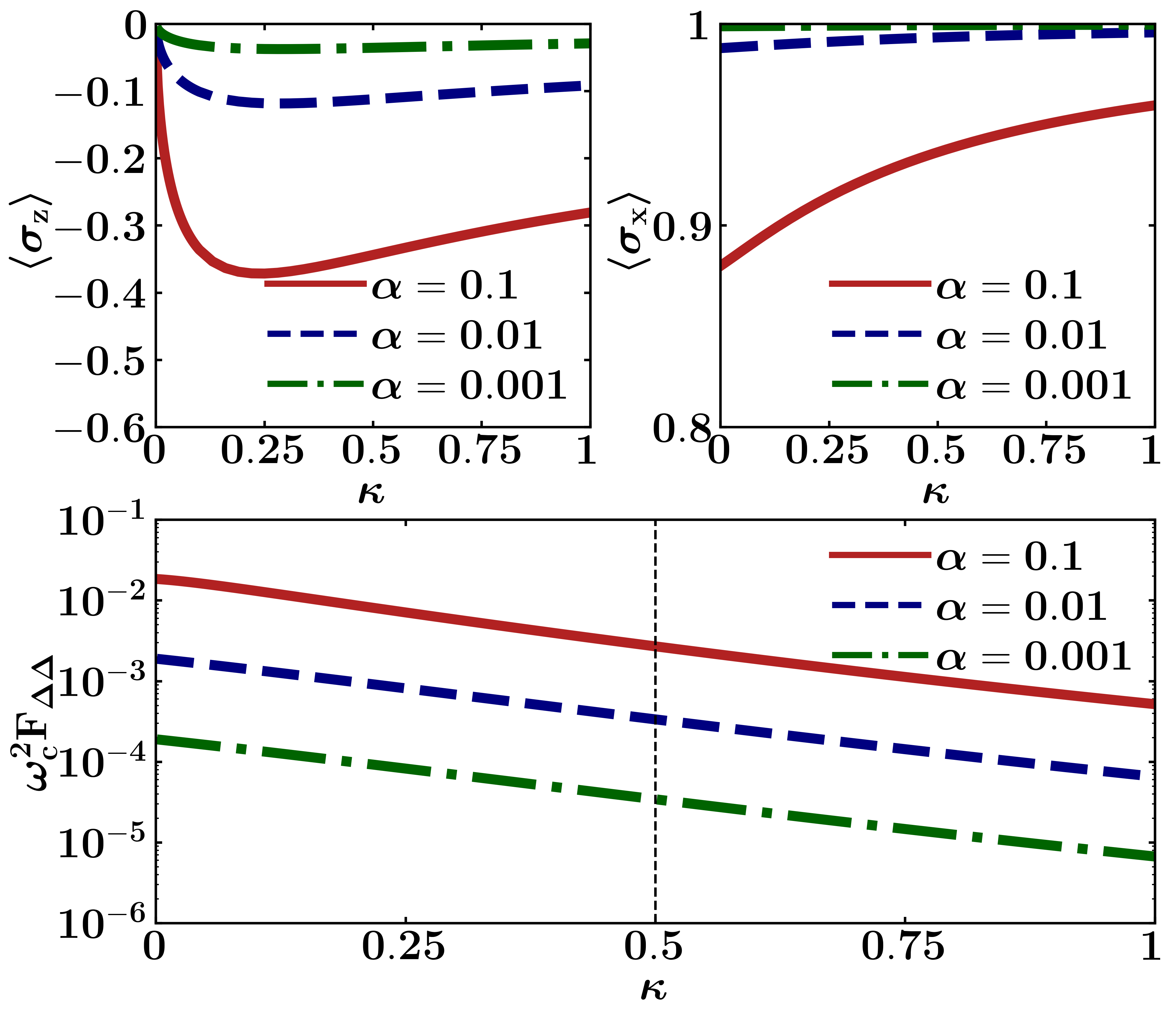}
    \caption{Ground state values of $\langle\sigma_z\rangle$ (top left), $\langle\sigma_x\rangle$ (top right) and of the QFI $F_{\Delta\Delta}$ (bottom) of the qubit while varying the dephasing coupling $\kappa$ for $T=0$, and three damping values $\alpha$. The vertical line at $\kappa=0.5$ indicates the crossover value: the stationary QFI at $\alpha=0$ diverges before the line and vanishes after it.}
    \label{Fig4}
\end{figure} 
We calculate the GS using the DMRG method. The average value of $\sigma_y$ is zero, $\langle\sigma_y\rangle=0$, because of a symmetry of $H$, while we plot $\langle\sigma_x\rangle$, $\langle\sigma_z\rangle$ and the QFI $F_{\Delta\Delta}$ in Fig.~(\ref{Fig4}).
As shown in Fig.~(\ref{Fig4}) the QFI does not exhibit a crossover at $\kappa=1/2$ in the static regime; the presence of the damping thus eliminates the peculiar behavior of the QFI, which we then assume can only be observed when the equilibrium state does not coincide with the ground state. This result, however, leaves open the question of the dynamical analysis, since also for pure dephasing at nonzero temperature the divergent behavior at long times disappears and is replaced by a crossover in the maximum value. As $\alpha\rightarrow0$ the QFI recovers the limit $F_{\Delta\Delta}\rightarrow0$ for any $\kappa$, while for $\alpha,\kappa\rightarrow0$ $F_{\Delta\Delta}\rightarrow+\infty$ since the GS tends towards a pure state, leading to divergent QFI, as can be seen from Eq.~\eqref{Eq3} in the limit $det(\rho)\rightarrow0$.

\subsection{Dynamics}\label{subsec:dynamics}
\begin{figure}
    \centering
    \includegraphics[width=1\linewidth]{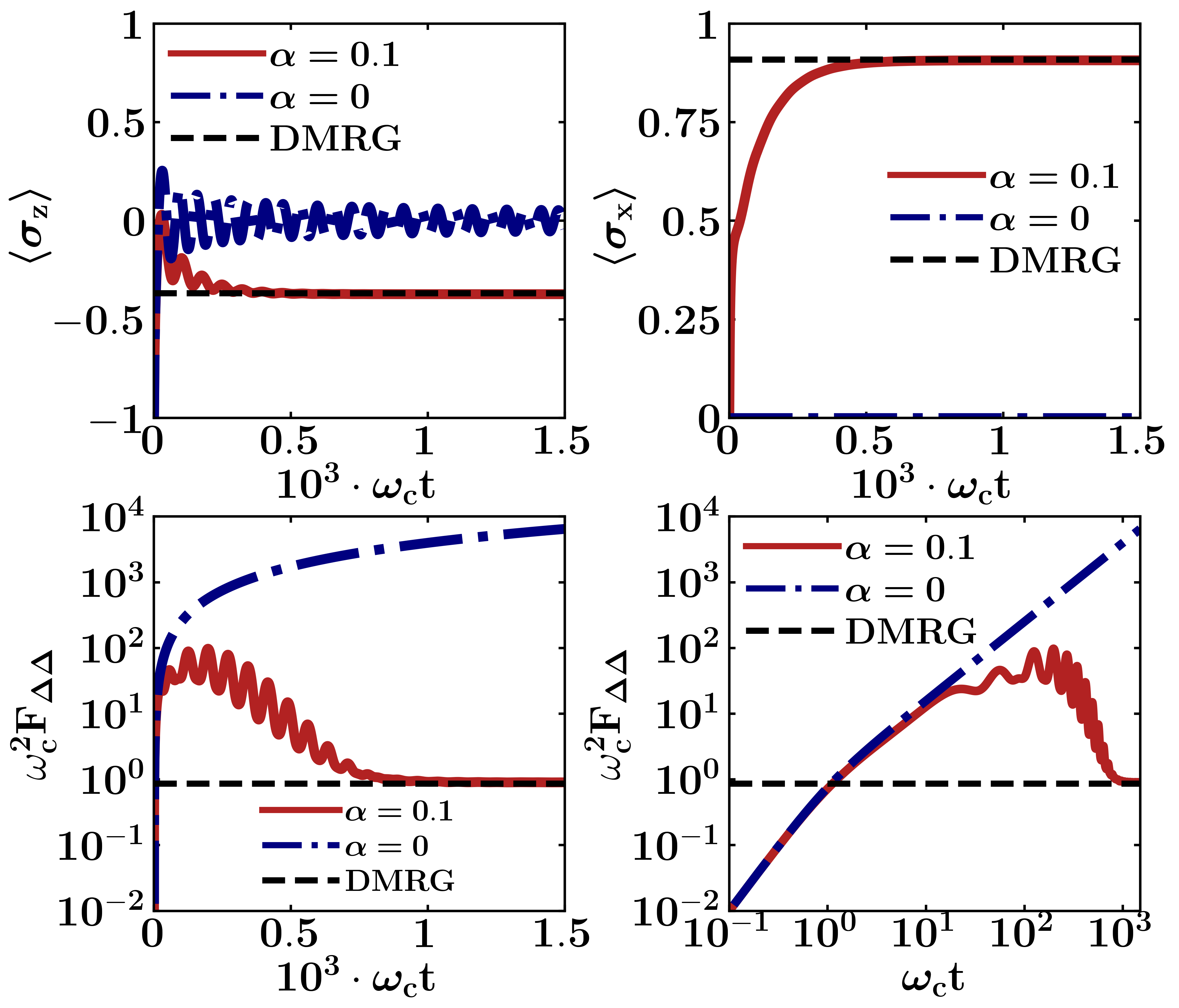}
    \caption{Value of $\langle\sigma_z\rangle$ (top left), $\langle\sigma_x\rangle$ (top right) and of the QFI $F_{\Delta\Delta}$ (bottom) of the qubit as a function of time for $T=0$, $\kappa=0.2$, and two values of $\alpha$: \(\alpha=0.1\) and \(\alpha=0\). The QFI is shown on both linear and logarithmic time scales. The {dashed} line provides the DMRG ground state value at $\alpha=0.1$ and $\kappa=0.2$.}
    \label{Fig5}
\end{figure}

\begin{figure}
    \centering
    \includegraphics[width=1\linewidth]{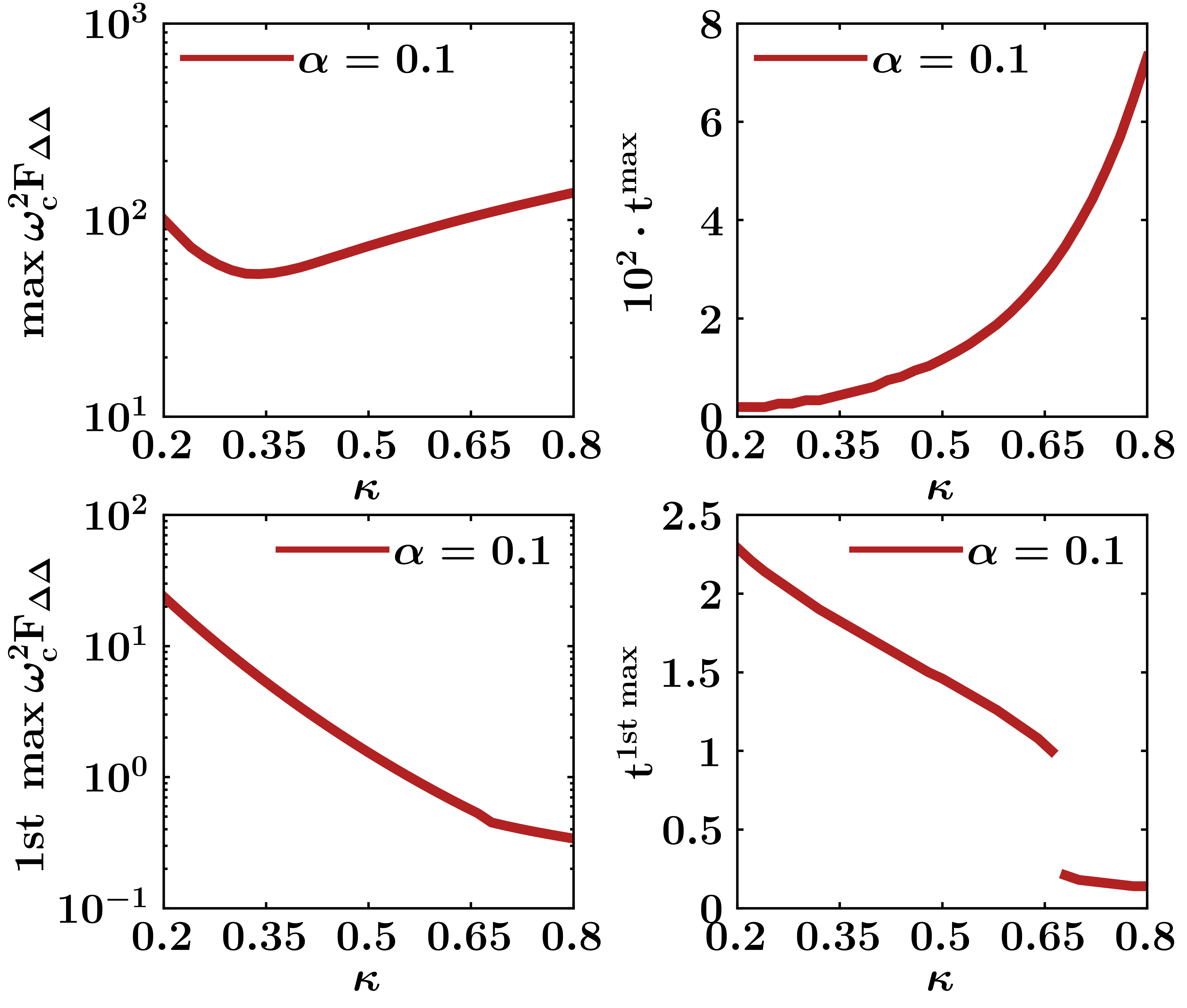}
    \caption{Value of the time maximum QFI $F_{\Delta\Delta}$ (top left), the time of maximum $t^{\text{max}}$ (top right), the QFI of the first local maximum of $F_{\Delta\Delta}$ (bottom left) and the time of the first maximum $t^{\text{1st max}}$ {(bottom right)} as a function of the dephasing coupling $\kappa$ for a damping coupling $\alpha=0.1$.}
    \label{Fig5b}
\end{figure}

Here we analyze the dynamical behavior of the QFI when turning on the amplitude-damping coupling.
 Using numerical simulations, we start from a qubit with gap $\Delta=\omega_c/10$ in the down state $\langle\sigma_z\rangle=-1$ and a vacuum bath state, and evolve the system for two values of {$\alpha$ while keeping $\kappa=0.2$ fixed.} The presence of amplitude-damping drives the system towards the ground state. The mean values $\langle\sigma_z \rangle$ and $\langle\sigma_y \rangle$ oscillate with frequency $\omega\approx\Delta$; we report $\langle\sigma_z\rangle$ in the upper left panel of Fig.~(\ref{Fig5}). The average value $\langle\sigma_x \rangle$, which was a conserved quantity in the pure-dephasing case, now exhibits nontrivial dynamics, as shown in the upper right panel of Fig. (\ref{Fig5}): after an initial increase, it temporarily stabilizes at a value that grows with $\alpha$. During this regime the dynamics slows down, until the effect of damping eventually prevails and $\langle\sigma_x\rangle$ reaches its GS value. The stronger the dephasing, the longer this plateau lasts. This behavior is consistent with the fact that the relaxation time diverges as the system approaches the integrable limit (here $\kappa/\alpha\rightarrow\infty$) starting from the nonintegrable model \cite{PhysRevResearch.7.023149}.
This effect cannot be attributed to critical slowing down, since we are far from the critical point $\alpha\ll1$. At short times the effect of the damping is small ($\partial_\Delta\ \langle\sigma_x\rangle\approx0$), so the Fisher information behaves similarly to the $\alpha=0$ case (bottom panels of Fig. (\ref{Fig5})), although the crossover shifts to a different value. As time increases, the term $\partial_\Delta\ \langle\sigma_x\rangle$ becomes nonnegligible, giving rise to a peak. Eventually, the derivatives $\partial_\Delta\ \langle\sigma_x\rangle$ and $\partial_\Delta\ \langle\sigma_z\rangle$ tend to constant values, causing the QFI to converge to its DMRG value. Thus, the presence of $\alpha\neq0$ breaks the crossover: it survives at short times $t$, but at long times $F_{\Delta\Delta}$ always converges to a finite value.
 \\ \\
We now address the question asked in the previous section: whether the crossover appears in the maximum value. We plot the maximum QFI varying $\kappa$ in the first panel of Fig.~(\ref{Fig5b}). As shown, the crossover is absent from this curve, which instead exhibits a smooth dependence on the coupling strength. An interesting feature is its nonmonotonic behavior: the maximum first decreases, then starts to grow. This can be explained by the derivatives' behavior: in particular, while the maximum of $\partial_\Delta\langle\sigma_x\rangle$ only grows with $\kappa$, that of $\partial_\Delta\langle\sigma_z\rangle$ decreases in absolute value, then grows again, giving rise to this behavior of the QFI.
The crossover can be identified by examining the early-time dynamics, specifically the first local maximum (bottom panels of Fig.~(\ref{Fig5b})), which displays similar behavior to the pure-dephasing, finite-temperature case of Fig.~(\ref{Fig2}). As $\kappa$ increases, the power-law dependence at small $t$ (note the logarithmic scale) exhibits progressively lower exponents, and at a value $\bar{\kappa}$ the local {maximum} undergoes an abrupt change, analogous to that observed in the pure-dephasing system. We observe this change in the time of the local maximum as well, as seen in the bottom right plot of Fig.~(\ref{Fig5b}). However, the presence of amplitude damping introduces a second characteristic timescale associated with relaxation, which depends primarily on the value $\alpha$ and determines the global maximum of the QFI. Consequently, the absolute maximum does not exhibit the crossover, as the dynamics in this regime are dominated by damping rather than by dephasing.
This also indicates that the damping caused by environment is reflected onto the order of the long-time limits. Indeed for $\alpha=0$ we have \begin{equation}\begin{cases}{cc}+\infty & \kappa<1/2\\ 1/\omega_c^2 & \kappa=1/2\\ 0 & \kappa>1/2\end{cases}=\underset{t\rightarrow+\infty}{\lim}{F_{\Delta\Delta}(\rho(t))}\neq F_{\Delta\Delta}(\underset{t\rightarrow+\infty}{\lim}{\rho(t)})=0,\end{equation}
where the second limit comes from the fact that $\partial_\Delta(\rho(+\infty))=0$. The relaxation introduced by damping restores the equality: \begin{equation}\underset{t\rightarrow+\infty}{\lim}{F_{\Delta\Delta}(\rho(t))}= F_{\Delta\Delta}(\underset{t\rightarrow+\infty}{\lim}{\rho(t)}).\end{equation}

\section{Conclusions}\label{sec:conclusions}
We have investigated the QFI associated with the estimation of the qubit energy splitting in the spin-boson model, focusing on the competition between coherent phase accumulation, pure dephasing, thermal fluctuations, and energy relaxation. In the exactly solvable pure-dephasing regime at zero temperature, we have analytically shown that the long-time behavior of the QFI is governed by a coupling-dependent algebraic scaling. As a consequence, a dynamical crossover occurs at the dephasing strength $\kappa=1/2$: for $\kappa<1/2$, the QFI grows without bound, although more slowly than in the isolated-qubit limit; at $\kappa=1/2$, it approaches a finite asymptotic value; and for $\kappa>1/2$, it vanishes at long times. This crossover originates from the competition between the coherent accumulation of the phase associated with the qubit gap and the progressive loss of coherence induced by the bosonic environment.

We have also studied the finite-temperature regime and shown that thermal fluctuations suppress the unbounded long-time growth of the QFI. The dynamics is then characterized by a finite-time maximum in the weak-coupling regime. Although the zero-temperature asymptotic crossover is removed, a clear signature of it remains in the coupling dependence of the maximum QFI, whose curvature changes around $\kappa=1/2$. Analytical expressions obtained in the weak- and strong-dephasing regimes, as well as close to the crossover, reproduce the zero-temperature behavior in the limit $T\rightarrow0$.

Furthermore, we have extended our analysis to non-Ohmic environments and identified a universal classification of dynamical gap sensing governed by the low-frequency bath exponent within this class of power-law environments. At zero temperature, sub-Ohmic environments suppress the asymptotic QFI, whereas super-Ohmic environments restore the quadratic scaling of coherent phase accumulation. The Ohmic case is therefore singular: it is the only regime in which the asymptotic QFI depends algebraically on the dephasing strength, giving rise to the coupling-controlled boundary at \(\kappa=1/2\). At finite temperature, thermal fluctuations modify this classification and shift the relevant boundary to the super-Ohmic regime \(s=2\). These results show that the low-frequency structure of the environment provides a universal control parameter for the long-time metrological behavior of the qubit.


Finally, at zero temperature we have introduced an additional weak amplitude-damping system-bath coupling, with strength $\alpha$, which enables population exchange and energy relaxation. This coupling breaks the conservation of the qubit energy that characterizes the pure-dephasing limit. In the weak-dephasing Ohmic regime, the short- and intermediate-time dynamics of the QFI retains a finite-time maximum and a visible signature of the pure-dephasing crossover. At long times, however, the behavior changes qualitatively: instead of diverging within the weak-coupling regime, the QFI approaches a finite value associated with the reduced interacting ground state. The damping channel therefore regularizes the unbounded sensitivity predicted in the ideal pure-dephasing model and sets a finite asymptotic precision bound.

These results are relevant for quantum sensing because they identify the conditions under which environmental dephasing does not necessarily suppress metrological sensitivity, but can instead coexist with an algebraically increasing QFI. At the same time, they show how finite temperature and energy relaxation introduce characteristic timescales that limit this enhancement. The present analysis thus provides a benchmark for distinguishing transient sensing advantages from genuinely asymptotic ones in structured open quantum systems.

Several extensions are worth pursuing. An important extension concerns systems of multiple qubits coupled either to independent reservoirs or to a common bosonic bath \cite{breuer2002theory,perroni2023first}. In that case, collective effects, bath-mediated correlations, and multipartite entanglement may modify both the scaling of the QFI and the position or even the nature of the dynamical crossover. Such developments would help assess whether the sensing enhancement identified in the single-qubit model can be amplified or stabilized in many-body quantum sensors. Additionally, the strong sensitivity of the QFI to environmental parameters points toward multiparameter quantum metrology. Investigating the joint estimation of the qubit energy and the coupling strength or bath temperature could reveal fundamental metrological trade-offs. Finally, the application of quantum optimal control techniques, such as continuous driving, may help to determine whether tailored control protocols can mitigate environmental noise for precise gap estimation.
\newpage 
\bibliography{main}

\newpage
\appendix
\section{Algorithm choice}\label{app:algorithm}
In this section we go into more detail about the numerical simulations. As mentioned above, we compared the results of the TEMPO algorithm to those obtained by simulating the joint dynamics of the environment and the qubit using the MPS and TDVP methods. To achieve convergence in TDVP simulations, we used $N=1000$ bosonic modes, computing the QFIM through Eq.~\eqref{Eq3}.
This comparison is worth clarifying, since in the previous work we relied exclusively on TDVP, raising the natural question of why we changed approach even though the system under study is very similar.
The first issue arises from the discrete implementation of the environment. As stated above, we model our bosonic bath as a set of $N$ harmonic oscillators, whose coupling constants $g_k$ must reproduce the continuous spectral density of the bath; a good discrete representation should approximately satisfy
\begin{equation}\sum_kg_k^2\approx \int_0^{+\infty}J(\omega)d\omega\;.\end{equation}
The first problem is therefore linked to the fact that, numerically, the spectral density $J_{\text{num}}(\omega)$ must necessarily have a finite numerical cutoff and does not extend to infinity: $J_{\text{num}}(\Omega,\omega)=J(\omega)\Theta(\Omega-\omega)$. In our case we choose a numerical cutoff $\Omega=4\omega_c$, a value that guarantees a good representation of the tail of $J(\omega)$. This, however, requires a very large number of modes, especially to obtain accurate long-time dynamics: with $N=1000$ we were able to reach a maximum time $t=2\pi/\Delta\omega\approx100$ with $\Delta\omega=(4\omega_c/N)$.
A second issue concerns the SVD (Singular Value Decomposition) cutoff value used in truncating the TDVP dynamics. The cutoffs typically used for high-precision simulations, on the order of $\sim10^{-10}-10^{-11}$, proved insufficient. The problem was not a slight inaccuracy in the final state, but rather the complete loss of part of the dynamics: at these cutoff values the system behaved as though it were subject only to pure dephasing, entirely ignoring the damping term, which we confirmed by the failure of the asymptotic state to thermalize. Probably this was caused by the fact that in the initial stages of the evolution, dephasing dominates and "occupies" all the available bond dimension, thereby preventing the damping contribution from being correctly represented. Using a cutoff $<10^{-12}$ we managed to obtain the correct observables (which now converged to the expected thermal values); however the long-time numerical stability of the QFI required an even lower cutoff, on the order of $\sim10^{-15}$. This inevitably caused the bond dimension to grow substantially; however, this alone does not fully account the extremely rapid bond dimension growth observed at early times, because, even with such a low cutoff, the system is in a weak damping regime. This is further supported by the fact that, for pure dephasing, simulations yield a maximum bond dimension of two, consistent with the exact solution; we would therefore expect a somewhat higher bond dimension in the weak-damping, strong-dephasing regime, but not such a sharp increase.
The reason for this lies in the fact that, initially, dephasing spreads correlations along the MPS chain; as a consequence, as the damping term becomes dominant it requires substantially more information, since each term must also carry the pre-existing dephasing correlations. This causes a sharp increase in bond dimension even though the actual entanglement entropy of the reduced state remains low.
For all these reasons, the MPS simulations turned out to be extremely slow, requiring up to a month for a single run. Moreover, as shown above, the times needed to reach the stationary state are quite long and increase further with the dephasing strength, making the MPS approach computationally unfeasible.
The TEMPO algorithm, implemented via the OQuPy library, resolves all of these issues. First, TEMPO incorporates the whole bath through the influence functional formalism, eliminating the discretization problem entirely, since the continuous, infinite spectrum is treated exactly. Moreover, the SVD truncation is only performed on the qubit, so the problem of the damping term being erroneously truncated out of the dynamics does not arise. Finally, and most importantly, the simulation time is dramatically reduced, since the bath dynamics need not be simulated numerically, but can instead be computed analytically via the influence functional. For the SBM in particular, the bath correlation function is known analytically, which further simplifies the calculations.


\end{document}